\documentclass[conference]{IEEEtran}

\usepackage{cite}
\usepackage{array}
\usepackage{amsfonts}
\usepackage{amsmath}
\usepackage{latexsym}
\usepackage{graphics}
\usepackage{amsbsy}
\usepackage{amssymb}
\usepackage{algorithm}
\usepackage{algorithmic}
\usepackage{enumerate}
\usepackage{color}
\usepackage{cite}
\usepackage{verbatim}
\usepackage{graphicx}%
\usepackage{multirow}
\usepackage{pifont}
\usepackage{algorithm}
\usepackage[table,xcdraw]{xcolor}

\begin{document}

\title{Spectral Attention-Driven Intelligent Target Signal Identification on a Wideband Spectrum}


\author{\IEEEauthorblockN{Gihan Mendis\IEEEauthorrefmark{1}, Jin Wei\IEEEauthorrefmark{1}, Arjuna Madanayake\IEEEauthorrefmark{2} and Soumyajit Mandal\IEEEauthorrefmark{3}}
\IEEEauthorblockA{\IEEEauthorrefmark{1}ECE Department\\
University of Akron \\
Akron, OH 44325\\
}
\IEEEauthorblockA{\IEEEauthorrefmark{2}ECE Department\\
Florida International University (FIU)\\
Miami, FL 33174 \\
}
\IEEEauthorblockA{\IEEEauthorrefmark{3}EECS Department\\
Case Western Reserve University\\
Cleveland, OH 44106\\
}
}

\maketitle

\begin{abstract}

This paper presents a spectral attention-driven reinforcement learning based intelligent method for effective and efficient detection of important signals in a wideband spectrum. 
In the work presented in this paper, it is assumed that the modulation technique used is available as a priori knowledge of the targeted important signal. 
The proposed spectral attention-driven intelligent method is consists of two main components, a spectral correlation function (SCF) based spectral visualization scheme and a spectral attention-driven reinforcement learning mechanism that adaptively selects the spectrum range and implements the intelligent signal detection. 
Simulations illustrate that the proposed method can achieve high accuracy of signal detection while observation of spectrum is limited to few ranges via effectively selecting the spectrum ranges to be observed. 
Furthermore, the proposed spectral attention-driven machine learning method can lead to an efficient adaptive intelligent spectrum sensor designs in cognitive radio (CR) receivers.

\end{abstract}


% \fi
\IEEEpeerreviewmaketitle
\section{Introduction}
Cognitive Radio (CR) was proposed as a solution for the problem of growing scarcity of electromagnetic spectrum for radio frequency (RF) communication and attracts interest in research community~\cite{mitola1999cognitive,mitola2009cognitive,haykin2005cognitive}.
Receivers with intelligent spectrum sensing capabilities are required for CR systems. Furthermore, detecting available modulation scheme on a carrier channel is a crucial step for spectrum sensing which is an essential function of CR systems~\cite{lee2008optimal,chou2007and}. Machine learning based methods are used in both wideband and narrowband spectrum sensing and other radio frequency signals related applications~\cite{raj2018spectrum,jiang2017machine,machuzak2016reinforcement}. In our previous work, we utilized spectral correlation function (SCF) based feature visualization along with deep learning methods such as deep belief networks (DBNs), convolutional neural network (CNNs) for automatic modulation classifications considering narrowband signals~\cite{mendis2019deep,mendis2018deep,mendis2017deep,mendis2016deep}. A major challenge for machine learning based wideband spectrum sensing applications in CR is that feature visualization method such as SCF on the wideband can be computationally expensive. But most of the features are unnecessary information for targeted application and therefore, only add an overload to increase the cost of CR receiver without adding value. Mechanisms are required that can adaptively identify the crucial ranges of the spectrum and the attention of further processing should be limited to the identified regions.

Attention-based machine learning methods are widely studied for applications in natural language processing (NLP) and machine vision. In~\cite{chorowski2015attention}, Chorowski \textit{et al.} proposed an end-to-end trainable speech recognition architecture based on an attention mechanism which combined both content and location information. In~\cite{bahdanau2016end}, Bahdanau \textit{et al.} proposed an end-to-end attention-based recurrent neural network (RNN) method for large vocabulary speech recognition. In~\cite{yin2016abcnn}, Yin \textit{et al.} proposed an attention-based convolutional neural network in company with a RNN for modeling sentence pairs.
In~\cite{mnih2014recurrent}, Mnih \textit{et al.} proposed an attention-based RNN model of visual attention, which was inspired by mechanisms of biological vision systems where visual information are selectively filtered according to an attention policy learned by past experience. 
In this work, we propose an attention-based RNN method for wideband spectrum sensing for CR. To the best of our knowledge, we are the first to propose attention-based deep learning method for applications in radio signal processing. 
This paper discusses a spectral attention-driven RNN based method for a wideband spectrum sensing application of target signal detection. We consider the modulation technique used as a know unique property of the target signal. SCF visualization is used on the spectrum ranges of attention to extract features useful for modulation detection.

The rest of the paper is organized as follows. The next section briefs the problem settings for the considered spectrum analysis and decision-making application of CR. Section~\ref{sec:method} shows the details of the proposed method including SCF-based feature visualization and spectral attention-driven detection mechanism. The simulation results and the conclusions are presented in Sections~\ref{sec:simulation} and~\ref{sec:conclusion}, respectively.

\section{Problem Setting}\label{sec:problem}
\begin{figure}[]
\center {\includegraphics[width=0.35\textwidth]{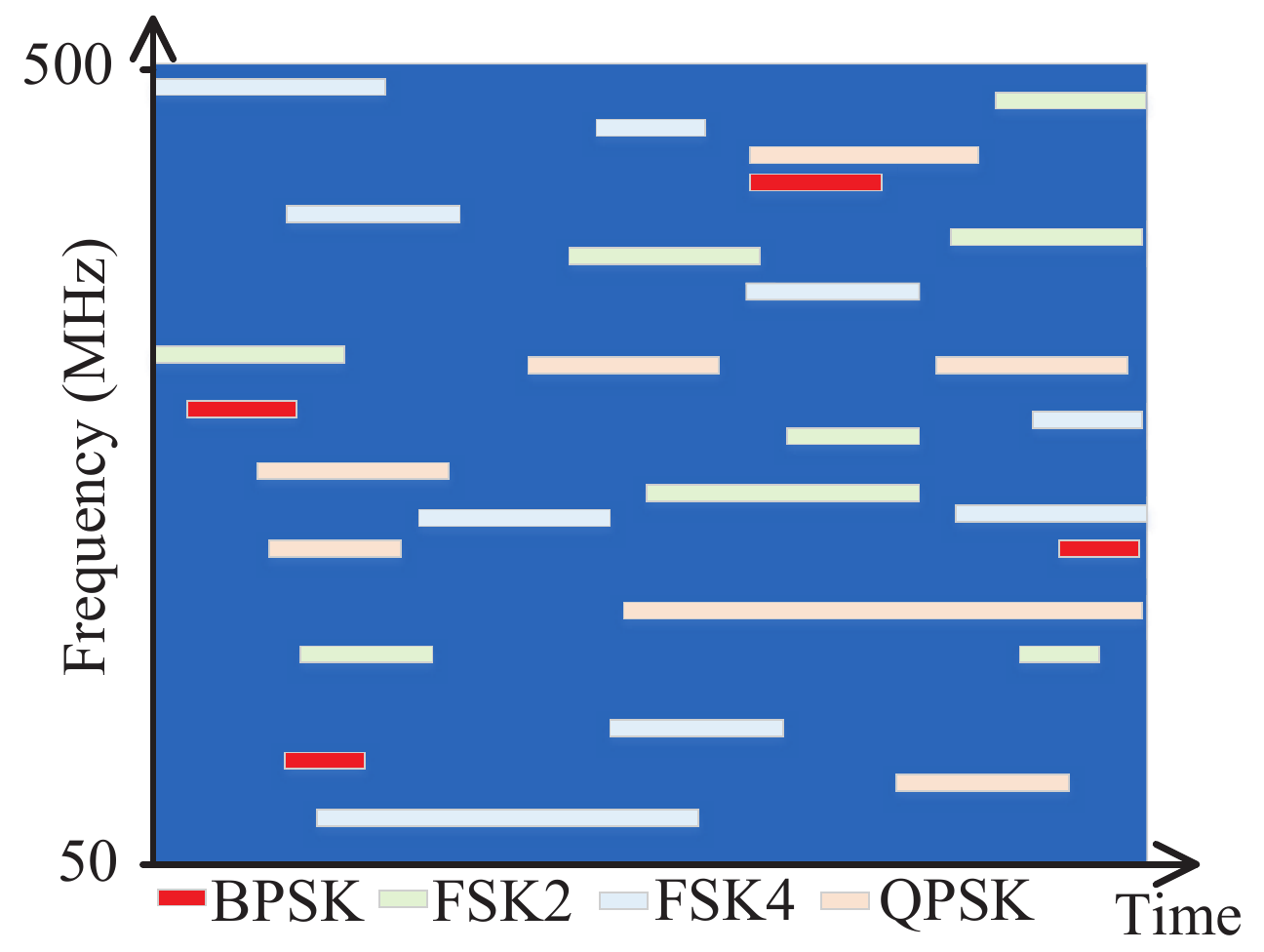}}
\caption{Spectrum of with target signal along with other signals.\label{Fig:Spectrum}}
\end{figure}

%In this paper, we investigate the idea of attention-based ML models on CR receiver. 
Our work focuses on achieving assured situational awareness on a wideband spectrum, which can congested, by developing a spectral attention-driven reinforcement learning-enabled intelligent scheme for target signal detection. As the first step in this research, in the work presented in this paper we assume that we have some priori knowledge of the target signal such as the modulation scheme. 
%
%The goal is to detect a target signal with some know properties in a wideband spectrum. In the simulation shown we consider a target signal with a known modulation scheme from a wideband spectrum with multiple background signals. 
While the receiver knows the modulation scheme of the target signal it does not know the carrier frequency. Additionally, there may exist frequency-hopping spread spectrum according to the situation of opportunistic channel selection of the CR transmitter. 
As an example, consider the frequency-time spectrum of a wideband background signal that contains FSK2, FSK4, and QPSK-modulated signals within a random bandwidth ranging between $50$ and $500$~MHz. A BPSK-modulated target signal is added at random locations of the frequency-time spectrum on top of the background, as shown in Fig.~\ref{Fig:Spectrum}. The objective of our work is to detect the presence of this target signal.

\section{Proposed Method}\label{sec:method}

\begin{figure}[]
\center {\includegraphics[width=0.35\textwidth]{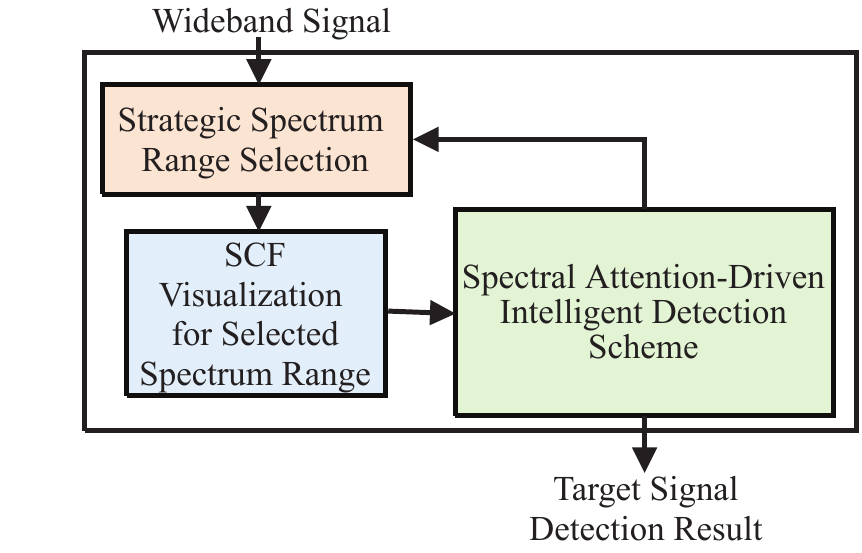}}
\caption{Overview of the proposed method.\label{Fig:Overview}}
\end{figure}

Figure~\ref{Fig:Overview} shows the overview of our proposed method. 
%
%Since the known property of the target signal is the modulation scheme, we consider the spectral correlation function (SCF) to characterize features associated with the modulation scheme of the observed signals.
Assuming some priori knowledge of the target signal, including the modulation scheme, is available, we employ the spectral correlation function (SCF) to pre-process the received RF signals and to visualize observed the spectrum environment. The output of our SCF-based visualization method is a 2-D ``image'' that characterizes the features associated with all the received signals. It is clear that this 2-D ``image" can has a large volume for the congested wideband spectrum analysis. Additionally, the features presented by the 2-D ``image" do not have equal values for contributing to the target signal detection. Therefore, to reduce the computational complexity and enable the real-time operation, we exploit reinforcement learning and deep learning techniques to develop a spectral attention-driven scheme that adaptively identify the critical features presented by the 2-D ``image" and selectively integrate these critical features to achieve the signal detection.  
Furthermore, the identified critical features presented by the SCF-enabled spectrum visualization implies the strategic selection of spectrum range for the signal detection, which can lead to the adaptive CR receiver design.

\subsection{SCF-based Feature Visualization}\label{sec:SCF}

Cyclic auto-correlation function~(CAF) is defined to quantize the amount of correlation between different frequency shifted versions of a given signal and represents the fundamental parameters of their second-order periodicity~\cite{}. Letting $T_0$ be the process period and $\alpha = \dfrac{\hat{m}}{T_0}$ be the cyclic frequency that indicates the cyclic evolution of the waveforms, where $\hat{m}$ is an integer, CAF can be calculated as following:
\begin{equation}
R^{\alpha}_x[l]= \left[\lim_{N\to\infty}\frac{1}{2N+1}\sum_{n=-N}^{N} x[n]x^*[n-l]e^{-j2\pi\alpha n}\right] e^{-j\pi\alpha l}
\label{eqscf3}
\end{equation}where $x[\cdot]$ denotes the modulated signal that is considered as cyclostationary process. Spectral correlation function~(SCF) is formulated by implementing Fourier transform on $R^{\alpha}_x$:
\begin{equation}
S^{\alpha}_x[f]= \sum_{l=-\infty}^{\infty} R^{\alpha}_x[l]  e^{-j2\pi fl}
\label{eqscf4}
\end{equation}where $R^{\alpha}_x[l]$ is defined in Eq.~(\ref{eqscf3}), and $f$ denotes the digital temporal frequency of the modulated signal. 
2-D SCF pattern can be obtained by calculating Eq.~(\ref{eqscf4}) for different values of $\alpha$ and $f$. For practical implementations, it is impossible to consider infinite number of samples, and thus we use considerably large number of samples. In our current implementations, time samples in a time window $N$ in Eq.~(\ref{eqscf3}) is limited to 25600 
% $l$ is consider to be integers in the range of [-12800,12800],
and $f$ in Eq.~(\ref{eqscf4}) is the digital frequency in the range of [$-\pi$,$\pi$] with the resolution of $pi/32$.

\begin{figure}[]
	 			\centering
	 			%\vskip -1ex
	 			\includegraphics[width=0.35\textwidth]{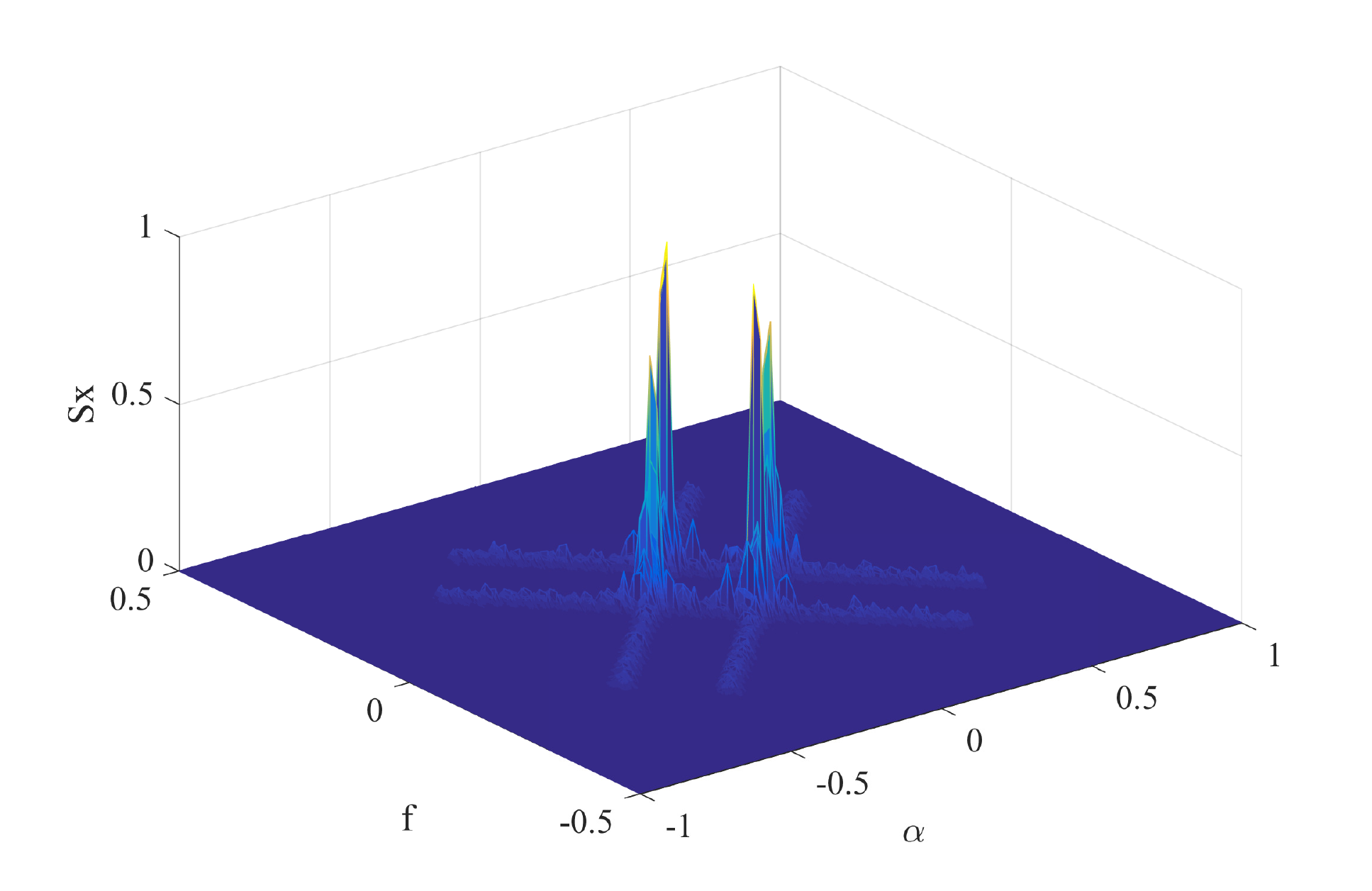}
	 			\caption{SCF pattern of simulated BPSK modulated signal (SNR of the simulated signal is 5~dB).}
                \label{Fig:BPSK3D}
			\end{figure}

\begin{figure}[]
	 			\centering
	 			%\vskip -10ex
	 			\includegraphics[width=0.35\textwidth]{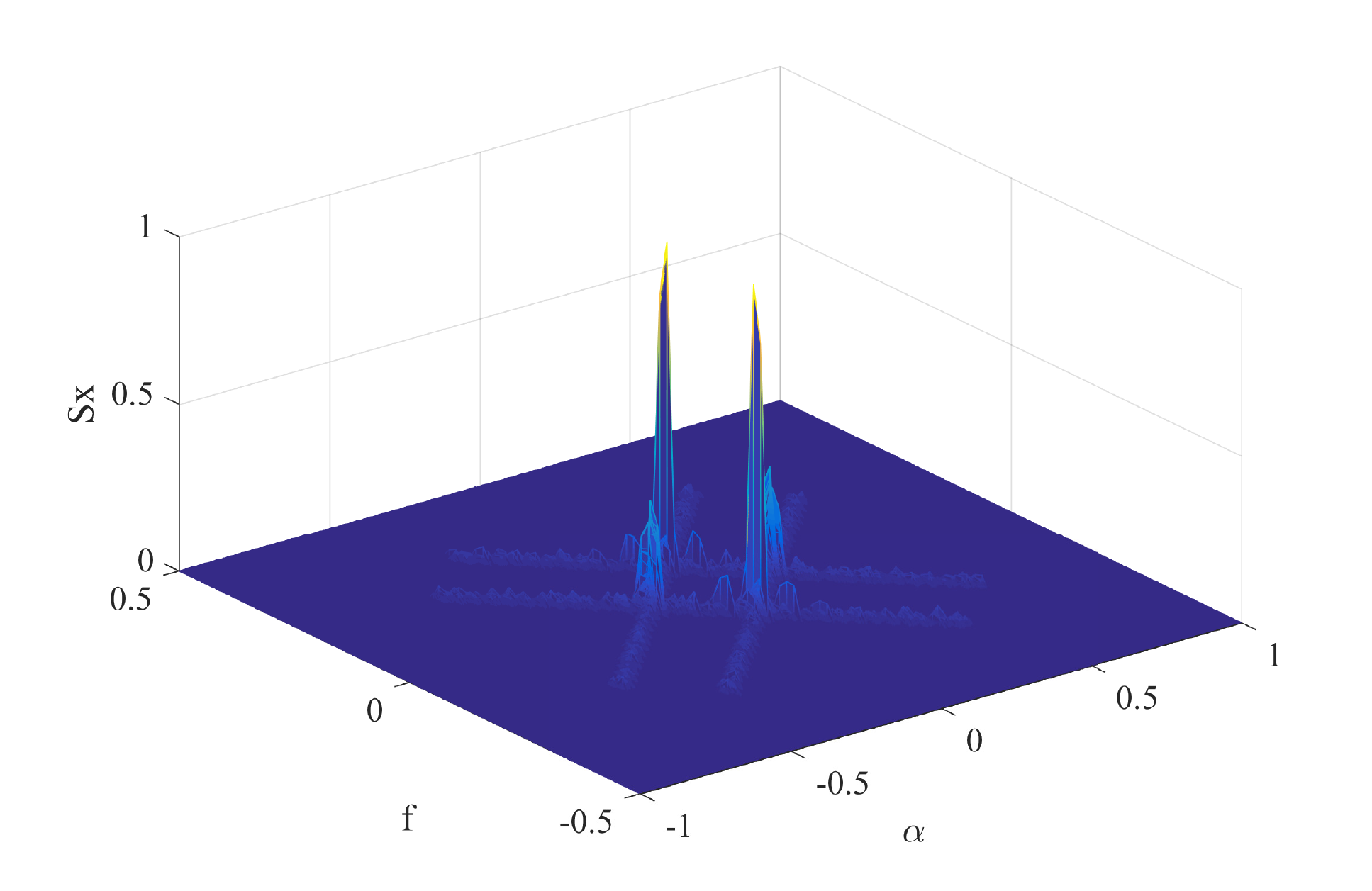}
	 			\caption{SCF pattern of simulated QPSK modulated signal (SNR of the simulated signal is 5~dB).}
                \label{Fig:QPSK3D}
			\end{figure}

\begin{figure}[]
	 			\centering
	 			%\vskip -5ex
	 			\includegraphics[width=0.35\textwidth]{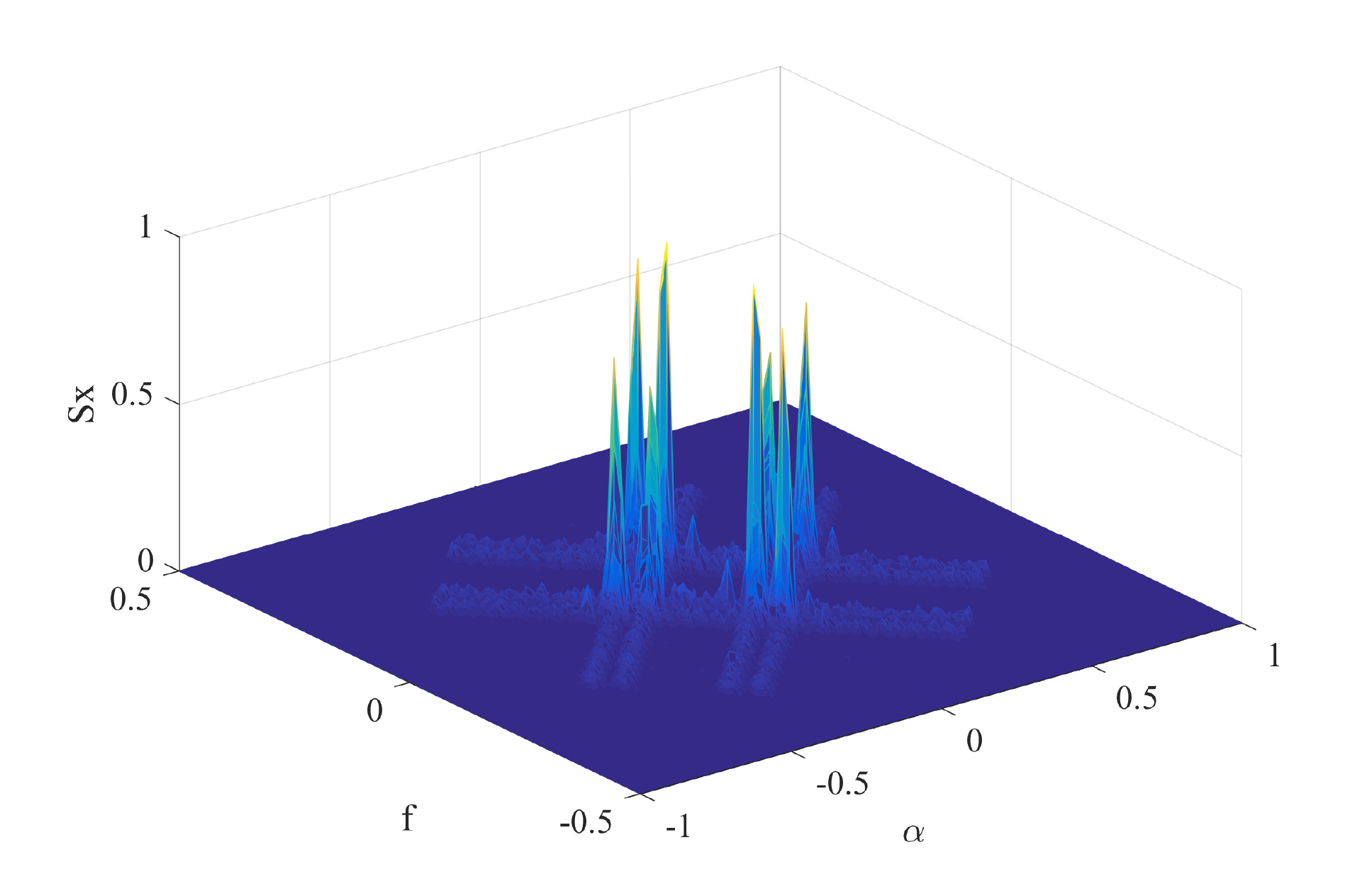}
	 			\caption{SCF pattern of simulated 2-FSK modulated signal (SNR of the simulated signal is 5~dB).}
                \label{Fig:2-FSK3D}
			\end{figure}

\begin{figure}[]
	 			\centering
	 			%\vskip -10ex
	 			\includegraphics[width=0.35\textwidth]{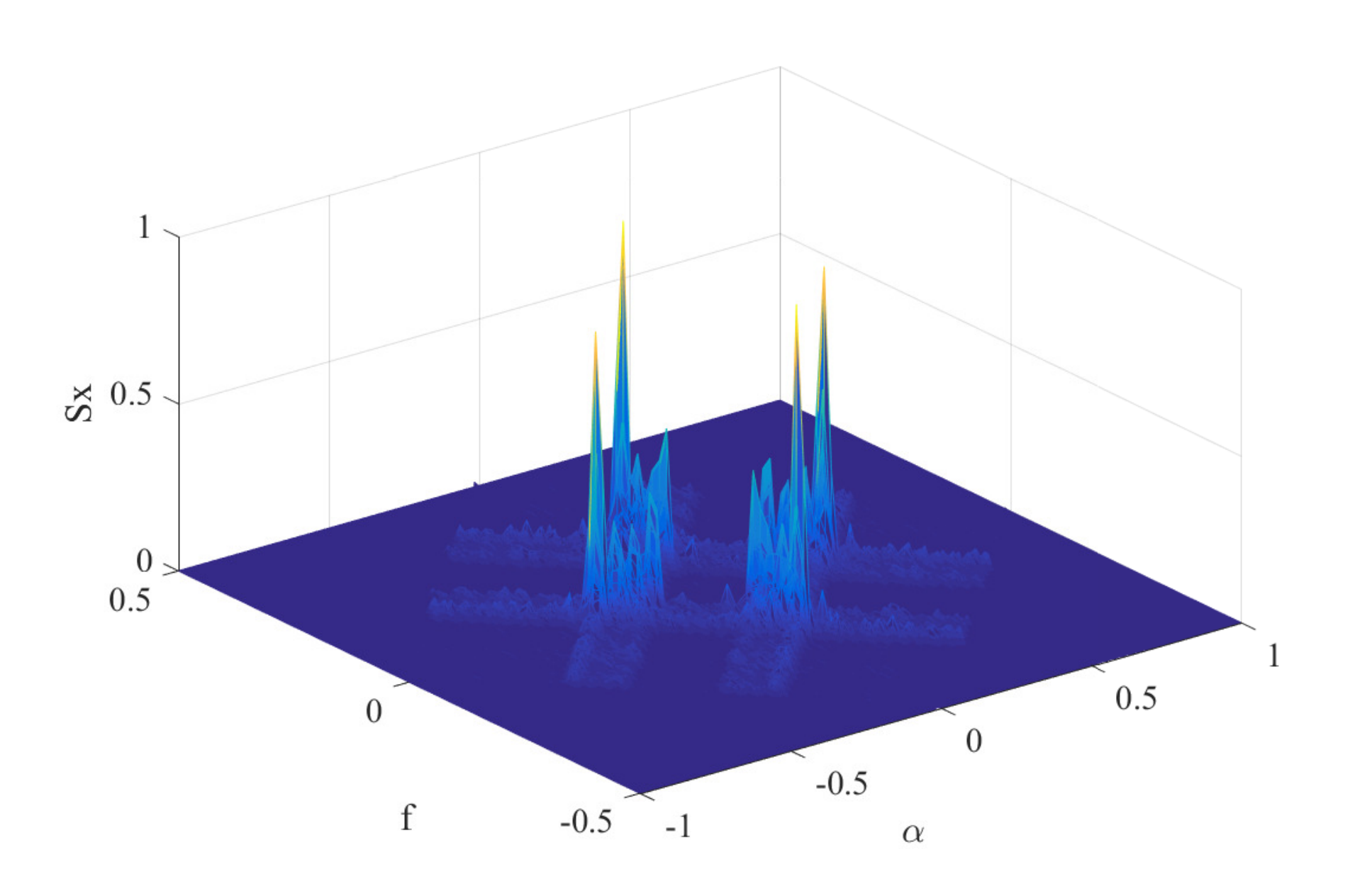}
	 			\caption{SCF pattern of simulated 4-FSK modulated signal (SNR of the simulated signal is 5~dB).}
                \label{Fig:4-FSK3D}
			\end{figure}

\begin{figure}[]
	 			\centering
	 			%\vskip -10ex
	 			\includegraphics[width=0.5\textwidth]{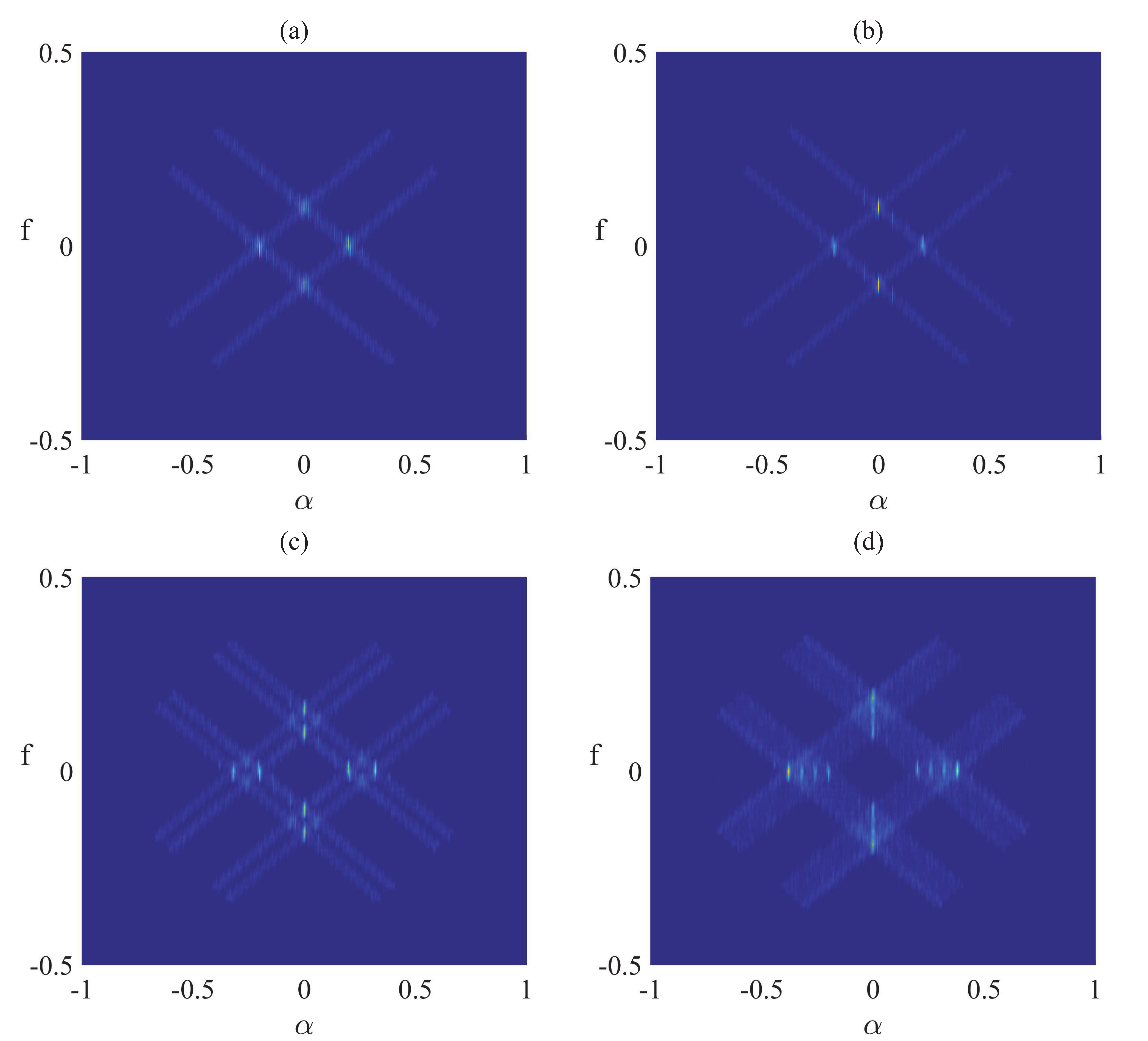}
	 			\caption{Top view of the SCF patterns of (a) BPSK; (b) QPSK; (c) 2-FSK; and (d) 4-FSK modulation schemes.}
                \label{Fig:SCF2D}
			\end{figure}

The SCF patterns effectively embody the features of the associated cyclostationary properties even in the presence of high additive white Gaussian noise (AWGN). This is because AWGN is a stationary process, and thus there are no
cyclostationary features in AWGN~\cite{gardner2006cyclostationarity}. Figures~\ref{Fig:BPSK3D}-\ref{Fig:4-FSK3D} present the SCF patterns obtained for modulated signals with 5~dB SNR for BPSK, QPSK, 2FSK, and 4FSK modulation schemes, respectively. Figure~\ref{Fig:SCF2D} shows the top-views of the above SCF patterns, in which lighter color intensity represent higher value.

\subsection{Spectral Attention-Driven Detection Mechanism}
\begin{figure}[]
	 			\centering
	 			%\vskip -10ex
	 			\includegraphics[width=0.35\textwidth]{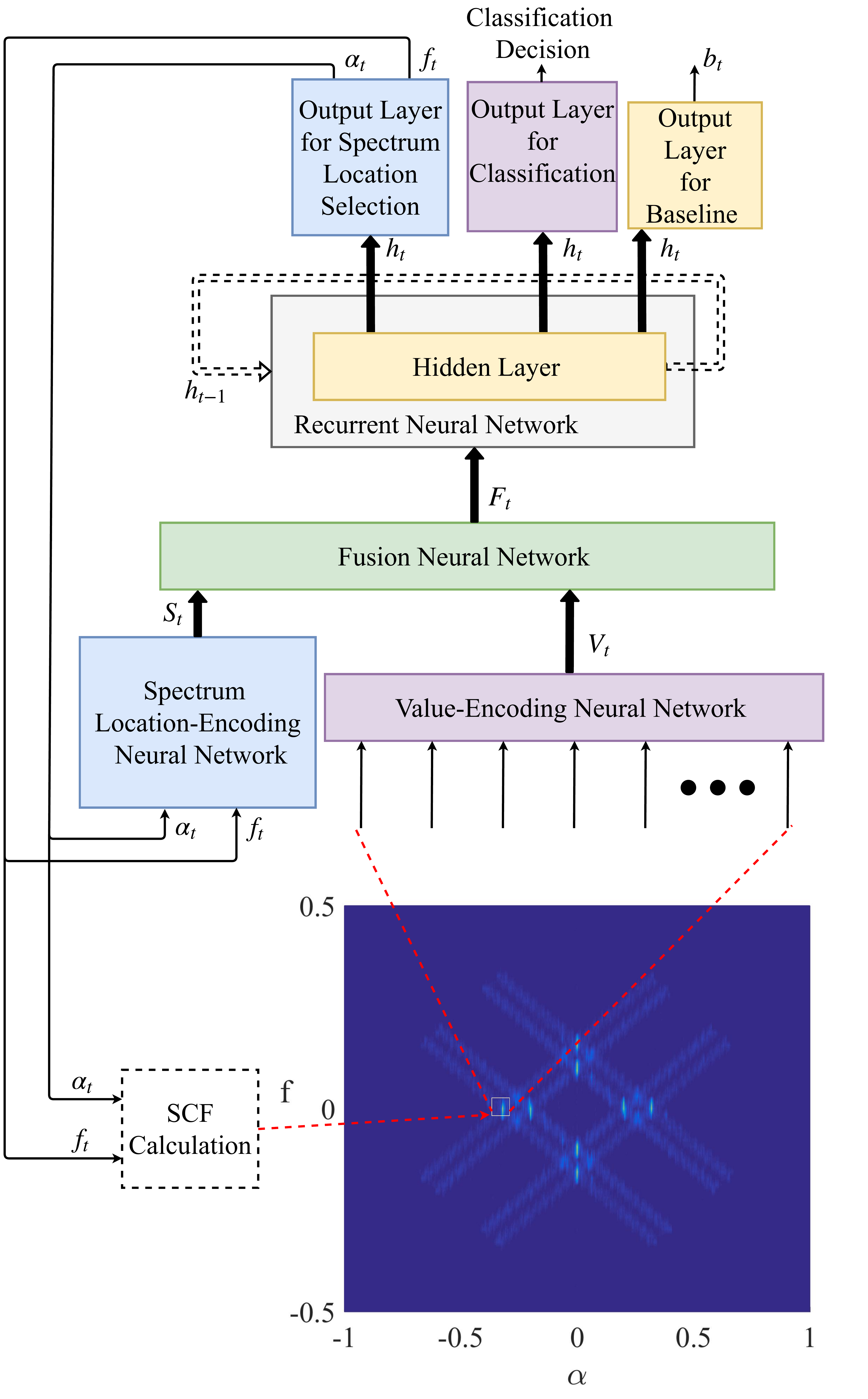}
	 			\caption{Reinforcement learning based spectral attention-driven method.}
                \label{Fig:RAM}
			\end{figure}

We exploit the reinforcement learning and deep learning techniques to develop a spectral attention-driven intelligent mechanism for detecting important signals on wideband spectrum. The overall structure of our mechanism is illustrated in Fig.~\ref{Fig:RAM}. 
The signals received according to the selected spectrum are visualized via our SCF-based method. As illustrated in Fig.~\ref{Fig:RAM}, the output of the SCF-based method is only a small patch, which is defined by the center temporal frequency $f_t$ and center cyclic frequency $\alpha_t$, of the 2-D ``image" that is potentially generated using the signals received across the whole wideband spectrum. The selected spectrum is called spectral attention in our work and the decision on the spectral attention is made adaptively by using our spectral attention-driven method.
%At time $t$, SCF for a selected range with fixed bandwidth around a given spectrum location is calculated as shown in Fig.~\ref{Fig:RAM}. The spectrum location is defined by center temporal frequency $f_t$ and center cyclic frequency $\alpha_t$. 
%
In our spectral attention-driven method, the SCF-based visualization output of the signals observed in the previously selected spectrum is reshaped as a vector and fed into the value-encoding neural network that generates the encoded representation for value $V_t$. Spectrum location-encoding neural network generates the encoded representation $S_t$ for the spectrum location $\{f_t, \alpha_t \}$. 
The fusion neural network is used to generate a fused representation $F_t$ of $V_t$ and $S_t$. A recurrent neural network~(RNN) is designed to characterize temporal features embedded in $F_t$. In our current design, the RNN has one hidden layer that is denoted by $h_t$ for time $t$. The hidden layer is generates via a neural network structure using the previous step value of hidden layer $h_{t-1}$ and $F_t$ as inputs. Additionally, the hidden layer $h_t$ is considered as the input for $3$ output layers for spectrum location selection, classification decision, and baseline $b_t$, which generate the spectrum location for the next step, binary detection decision, and baseline $b_t$ for reinforcement learning based training of the neural network structure, respectively. Spectrum location and classification decision are sampled for $T$ steps.      

The neural network structure of the proposed spectral attention-driven method is trained using a policy gradient reinforcement learning (RL) based method. 
Two main components of RL are the agent and the environment.
Main parameters that define these components are, states, actions, and rewards. The environment of the RL model can be in different states. The environment changes its state according to the action performed by the agent. The agent performes action according to a policy and receives a reward according to the state of the environment or according to the (state, action) pair. The policy of the agent creates a probapility distribution over posible interation sequences. The goal of RL is to find the optimum policy that maximizes the expectation of the total reward under this distribution~\cite{sutton2018reinforcement}. In policy gradient based RL methods, this is archived by iteratively updating the parameter that defines the policy.
In our work, the main parameters of the RL are modeled as follows,

\textbf{States:} The hidden layer of the RNN $h_t$ summarizes the information extracted from the history of past observations. We model the state of the hidden layer $h_t$ as the state of the RL. 

\textbf{Actions:} Two parameters of the proposed method are considered as actions in the RL model. Those are, the spectrum location $\{f_t, \alpha_t \}$ and the binary detection decision outputs of the neural network structure.

\textbf{Rewards:} In our work, classification decision make after $T$ steps is considered as important. In order to model this the cumulative reward $R$ is calculated as $R= \sum_{t=1}^{T} r_t$, where $r_t=1$ if classification decision is correct at $t=T$ and $r_t = 0$ otherwise. 

The total expected reward is given as following,
\begin{equation}
J(\theta)= \mathbb{E}_{\pi_{\theta}(s_{1:T})}[R]
\label{expected_reward}
\end{equation}where $\pi_{\theta}$ is the policy of the agent and $\theta$ represents policy parameters, which includes all the weight and bias variables of the neural network structure in Fig.~\ref{Fig:RAM}.

Considering the problem as a partially observable Markov decision process (POMDP), the gradient of the total expected reward can be approximated as following,
\begin{equation}
\nabla_{\theta}J(\theta)\approx \dfrac{1}{M}\sum_{i=1}^{M}\sum_{t=1}^{T}\nabla_{\theta}log \pi_{\theta}(a_t|s_{1:T})(R_t^{i} - b_t)
\label{gradient}
\end{equation} where $\pi_{\theta}$ is the current policy with parameter $\theta$, $b_t$ is a baseline obtained from the baseline output of the neural network structure in Fig.~\ref{Fig:RAM}, and $M$ is Monte Carlo sampling number.

A loss function $L$ is defined considering reward maximization, regularization of above parameter $b_t$ and reducing classification error. The gradient of this loss function $L$ is calculated as following,
\begin{equation}
\nabla_{\theta}L = -\nabla_{\theta}J(\theta) + \nabla_{\theta}\{ \dfrac{1}{M T}\sum_{i=1}^{M}\sum_{t=1}^{T}(R_t^{i} - b_t)^2 + \dfrac{1}{M}\sum_{i=1}^{M}(y - \bar{y}) \}
\label{Loss}
\end{equation}where $y$ is the actual label and $\bar{y}$ is the predicted label in the $T$th step.
Using the gradient calculated according to Eq.~(\ref{Loss}), the parameters $\theta$ are updated iteratively with following gradient descent update rule,
\begin{equation}
\theta_{n+1} = \theta_{n} - \alpha \nabla_{\theta}L
\label{update_rule}
\end{equation}where $\alpha$ is the learning rate and n is the index of the training trial.
%Spectrum location and classification decision are sampled for $T$ time-steps. Cumulative reward $R$ is calculated as $R= \sum_{t=1}^{T} r_t$, where $r_t=1$ if classification decision is correct at $t=T$ and $r_t = 0$ otherwise. 
%
%
After training, the network is expected to learn a policy to decide what $T$ locations on the spectrum to look at to make a reliable classification decision at the step $T$. This will reduce the computation cost of calculating SCF on a wide bandwidth.

\section{Simulations and Results}\label{sec:simulation}
In this section, we will evaluate our proposed spectral attention-driven signal detection mechanism by considering two scenarios. The parameters of our mechanism used for the simulations are stated in Table~\ref{Tab:Parameters}.  

\begin{table}[]
\caption{Parameters of the spectral attention-driven methods used in the simulations\label{Tab:Parameters}}
\begin{tabular}{|c|c|}
\hline
\rowcolor[HTML]{CBCEFB} 
Parameter & Vector Dimension \\ \hline
Input & 16 \\ \hline
Steps $T$ & 5 \\ \hline
Encoding for value $V_t$ & 128 \\ \hline
Encoding for spectrum location $S_t$ & 128 \\ \hline
Fused representation $F_t$ & 256 \\ \hline
Hidden layer $h_t$ & 256 \\ \hline
Spectrum location output & 2 \\ \hline
Classification output & 1 \\ \hline
\end{tabular}
\end{table}

In both scenarios, we consider detecting the important signal within $500$~MHz spectrum that is divided into $64$ bins for implementing our spectral attention-driven mechanism.
%
%That is the bandwidth of the tunable bandpass filter simulated is approximately 46~MHz. 
%The number of time-steps ($T$) used in the simulations is $5$. 
Additionally, we assume the number of Steps ($T$) for achieving adaptive decision making on spectral attention is $5$ in both scenarios. 
In all the simulations, 800 received signals are simulated and used as training data. 200 signals are simulated and used as testing data. 
Furthermore, received signals are simulated by using MATLAB/Simulink, and spectral attention-driven machine learning method is implemented by using Python.
%Matlab/Simulink computation and simulation software are used for simulation of received signals and calculation of SCF. Recurrent attention method based machine learning is performed in Python programming language using TensorFlow machine learning tools.  

\begin{figure}[]
	 			\centering
	 			%\vskip -10ex
	 			\includegraphics[width=0.45\textwidth]{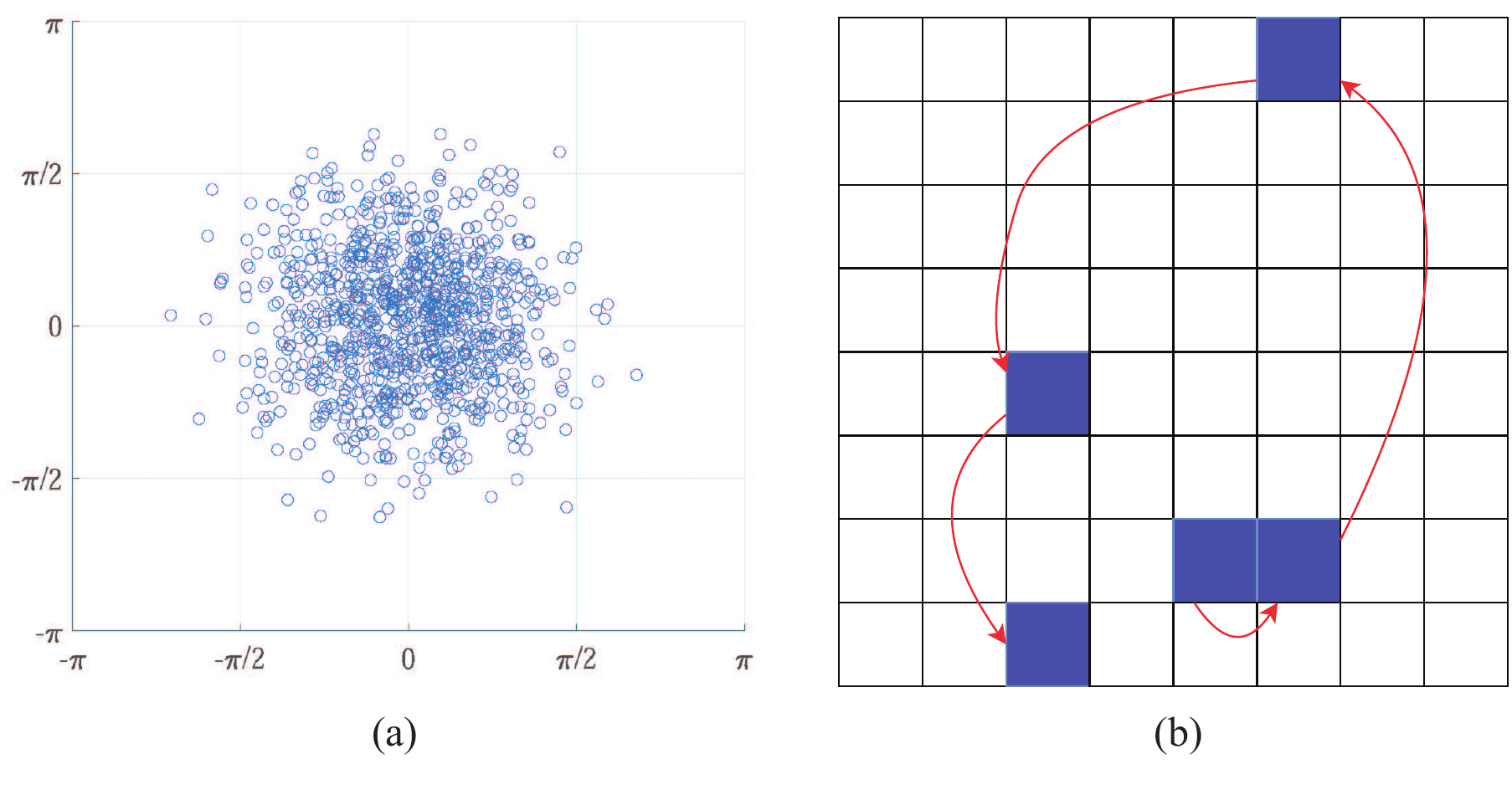}
	 			\caption{(a) Scatter plot of center frequencies of all considers spectral attentions for classifying 200 observed signals; (b) Example of the reduced SCF pattern calculation for the proposed 5 step spectral attention-driven method, where blue color indicates the spectral attentions that SCF calculation are required and the arrows show the transitions of attention for the 5 steps.}
                \label{Fig:grid}
			\end{figure}

Figure~\ref{Fig:grid}(a) shows the scatter plots of center frequencies of all spectral attentions considered in a classification task with $97\%$ accuracy that classifies 200 received signals with BPSK and 4FSK modulated signal on the 300~MHz carrier frequency. Spectral attention is randomly initialized with uniform distribution while training. The observed center frequencies of spectral attentions are loosely within the peaks of the SCF pattern. This verifies that the proposed method learns a strategy that considers the spectral properties of the observed signal.  
Figure~\ref{Fig:grid}(b) shows spectral attentions that require SCF calculation for a single trial of classification. The full grid represents the whole temporal and cyclic spectrum of SCF considered and the parts in blue are the spectral attentions where SCF values are used for the machine learning. Therefore, only $5/64$ of SCF is needed to for the proposed attention based method, which is a large reduction in computation cost. 

%
%Simulations are performed considering two scenarios. 
\subsection{Scenario I}
\begin{figure}[]
	 			\centering
	 			%\vskip -10ex
	 			\includegraphics[width=0.45\textwidth]{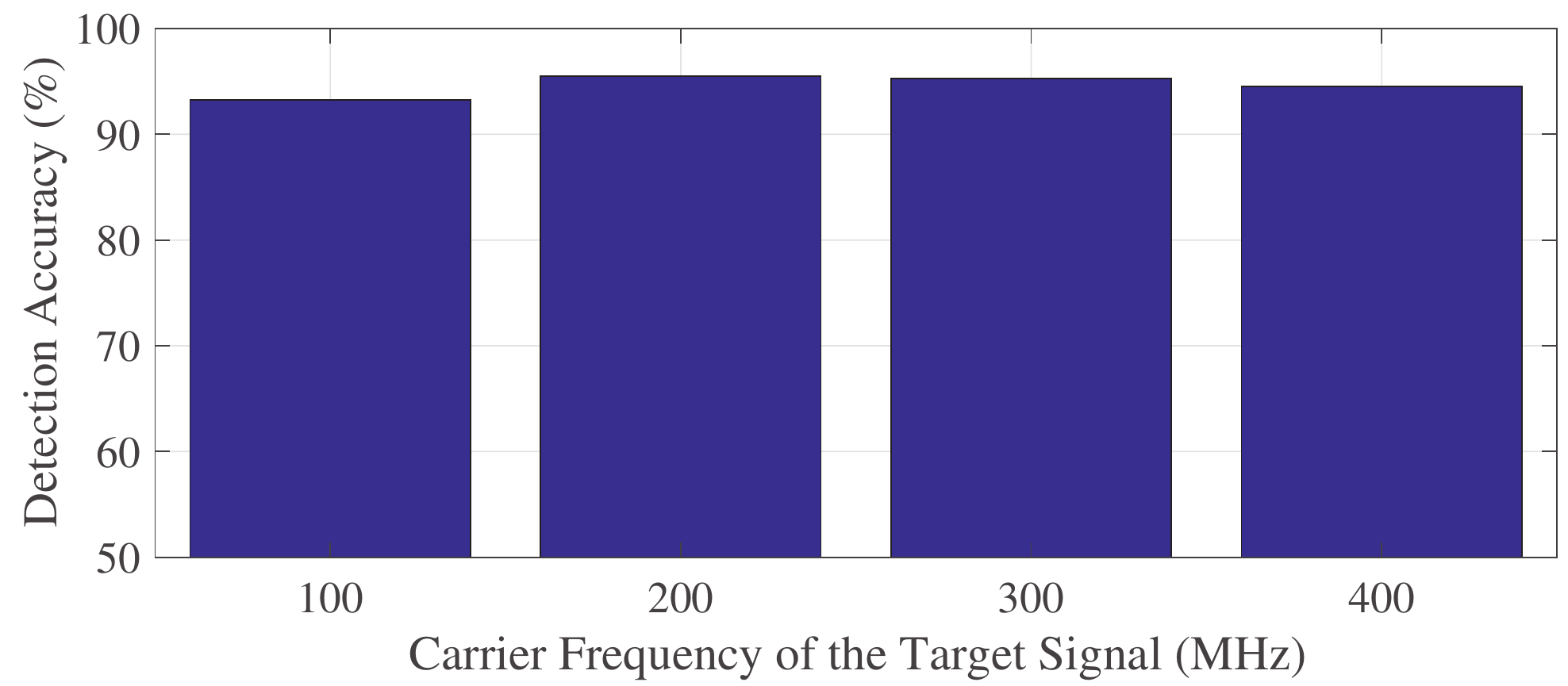}
	 			\caption{Detection accuracy of the target BPSK signal in Scenario I.}
                \label{Fig:sc1}
			\end{figure}

In this scenario, we assume that there is only one of the BPSK, QPSK, 2FSK, and 4FSK modulated signals is present at the receiver in the considered time window. The carrier of the modulated signal is randomly selected from 100~MHz, 200~MHz, 300~MHz, and 400~MHz. The goal is to detect the presence of the target BPSK modulated signal in the wideband spectrum of 0-500~MHz. Figure~\ref{Fig:sc1} shows the detection accuracies obtained from the proposed method when the target signal is in different carrier frequencies. For all considered carrier frequencies the detection accuracy remained above $93\%$ for this scenario.

\subsection{Scenario II}
\begin{figure}[]
	 			\centering
	 			%\vskip -10ex
	 			\includegraphics[width=0.45\textwidth]{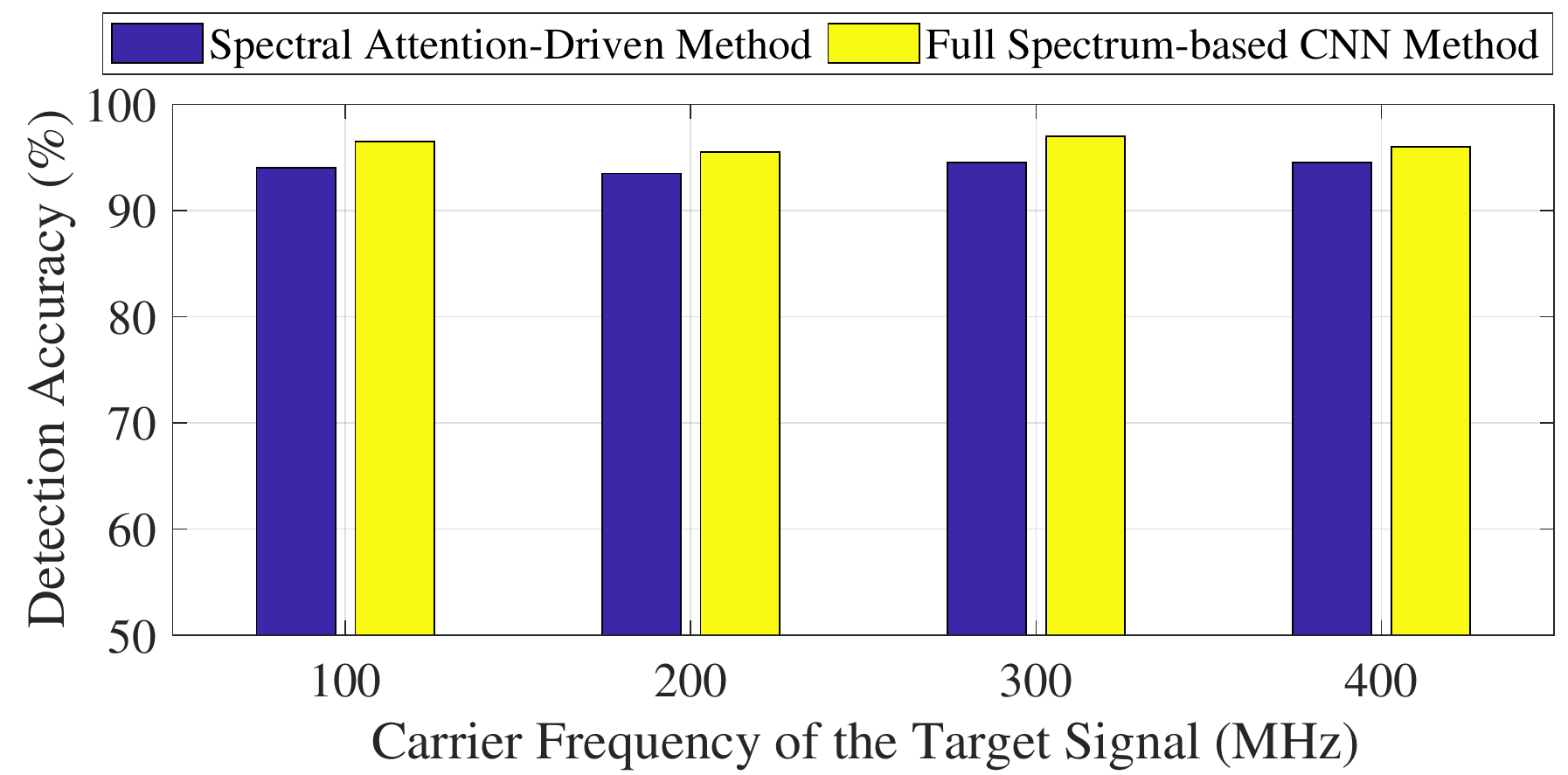}
	 			\caption{Detection accuracy of the target BPSK signal in Scenario II.}
                \label{Fig:sc2}
			\end{figure}

In the Scenario II, we consider a background with one, two, or none of QPSK, 2FSK, 4FSK modulated signals. The target BPSK modulated signals are present in some of the received time windows. Similar to Scenario I, carriers of the modulated signals, including target and background signals, are randomly selected from 100~MHz, 200~MHz, 300~MHz, and 400~MHz. The goal is to detect the presence of the target BPSK modulated signal in the wideband spectrum. Figure~\ref{Fig:sc2} compares the detection accuracies obtained from the proposed spectral attention-driven method and full spectrum-based convolutional neural network (CNN) classifier. The CNN used contains 3 convolution layers. 
From Fig.~\ref{Fig:sc2}, we can see that proposed method shows above $93\%$ accuracy for all considered frequencies. Also, it is clear that the accuracies obtained from the proposed method are comparable with the full spectrum-based CNN method that uses full SCF pattern as input.

\section{Conclusions}\label{sec:conclusion}
In this paper, we present our research on developing a spectral attention-driven method for effectively detecting the important signals in a wideband spectrum. As the first step in this research direction, in the work proposed in this paper, we assume that some priori knowledge of the target signals, including the modulation techniques, is available for the detection. Our spectral attention-driven method proposed in this paper consists of two main components, a SCF-based spectral visualization scheme and a spectral attention-driven mechanism that adaptively selects the spectrum range and implements the intelligent signal detection. As illustrated in the simulation, our proposed method can achieve high accuracy of signal detection via effectively selecting the spectrum range to be observed. Furthermore, with the spectral attention, our proposed method can lead to an adaptive CR receiver design.
\bibliographystyle{IEEEtran}
\bibliography{gihan_ref}

\end{document}